\documentclass{iau} 
\usepackage{graphicx}
\usepackage{color}

\title[JD 11.~~Neutron star crust cooling] 
{A window into the neutron star: \\ Modelling the cooling of accretion heated neutron star crusts}

\author[Wijngaarden \etal ]   
{Marcella J.P. Wijngaarden$^{1,2}$,
Rudy Wijnands$^2$, Laura S. Ootes$^2$, Aastha S. Parikh$^2$ \& Dany Page$^3$}

\affiliation{$^1$Mathematical Sciences, University of Southampton, \\ SO17 1BJ, Southampton, United Kingdom \\email: {\tt M.J.P.Wijngaarden@soton.ac.uk} \\[\affilskip]
$^2$Anton Pannekoek Institute for Astronomy, University of Amsterdam, \\ Postbus 94249,
NL-1090 GE, Amsterdam, the Netherlands
\\[\affilskip]
$^3$Instituto de Astronom\'iía, Universidad Nacional Aut\'onoma de M\'eéxico, \\ Mexico City, CDMX 04510, Mexico} 
\pubyear{2017}
\volume{337}  
\jname{Pulsar Astrophysics -­ The Next 50 Years}
\editors{P. Weltevrede, B.B.P. Perera, L. Levin Preston \& S. Sanidas, eds.}

\begin{document}

\maketitle

\begin{abstract}
In accreting neutron star X-ray transients, the neutron star crust can be substantially heated out of thermal equilibrium with the core during an accretion outburst. The observed subsequent cooling in
quiescence (when accretion has halted) offers a unique opportunity to study the structure and
thermal properties of the crust. Initially crust cooling modelling studies focussed on transient X-ray
binaries with prolonged accretion outbursts ($>$ 1 year) such that the crust would be significantly
heated for the cooling to be detectable. Here we present the results of applying a
theoretical model to the observed cooling curve after a short accretion outburst of only $\sim$10 weeks.
In our study we use the 2010 outburst of the transiently accreting 11 Hz X-ray pulsar in the globular
cluster Terzan 5. Observationally it was found that the crust in this source was still hot more than
4 years after the end of its short accretion outburst. From our modelling we found that such a
long-lived hot crust implies some unusual crustal properties such as a very low thermal conductivity
($>$ 10 times lower than determined for the other crust cooling sources). In addition, we present our
preliminary results of the modelling of the ongoing cooling of the neutron star in MXB 1659-298.
This transient X-ray source went back into quiescence in March 2017 after an accretion phase of
$\sim$1.8 years. We compare our predictions for the cooling curve after this outburst with the cooling curve of the same source obtained after its previous outburst which ended in 2001.
\end{abstract}

\firstsection 
\vspace*{-0.3 cm}
\section{Introduction}

The study of the thermal evolution of neutron stars can reveal important properties of the neutron star interior. A particular useful subclass for these kind of studies is formed by neutron stars residing in transient low-mass X-ray binaries (LMXBs). In these systems, the neutron star occasionally accretes matter from a companion star via Roche-lobe overflow. During such an accretion outburst, a series of nuclear reactions are induced in the crust which can significantly heat it up out of thermal equilibrium with the core (see, e.g., \cite[Haensel \& Zdunik 1990]{haensel1990} and \cite[Rutledge et al. 2002]{2002rutledge}). When the accretion outburst has halted, the subsequent cooling of the crust can be observed over a typical timespan of several years. This makes it possible to study the thermal evolution of the crust as it returns to thermal equilibrium with the core within a human lifetime (for a recent review of neutron star crust cooling we refer to \cite[Wijnands et al. 2017]{2017wijnands}). The cooling rate is set by the heat capacity and thermal conductivity properties of the crust, and thus allows us to probe the structure and composition of the neutron star crust (\cite[Page \& Reddy 2012]{page2012}). From previous crust cooling studies it was found that for multiple sources the observed surface temperatures were larger than could be explained with crustal heating models (e.g., \cite[Brown \& Cumming 2009]{Brown2009}). This showed that an additional heat source at shallow layers ($\sim$150 meters) in the crust must release a significant amount of heat (typical 1-2 MeV per accreted nucleon, although in one source it could be as much as $\sim$10 MeV/nucleon, \cite[Deibel et al. 2015]{deibel2015}, \cite[Parikh et al. 2017b]{2017parikhb}). The origin of shallow heating is unknown and remains one of the open questions in crust cooling studies. In this work, we present the implications of modelling the thermal evolution of two cases of neutron star LMXBs: Terzan 5 X-2 and MXB 1659-29. We make use of the most recent version of the theoretical cooling code \textit{NSCool} (\cite[Page 2016]{2016ascl}) as described in \cite[Ootes et al. (2016)]{2016lsootes}. 

\vspace*{-0.35 cm}
\section{Terzan 5 X-2: A low conductivity crust}

The second X-ray transient discovered in the globular cluster Terzan 5 was found to be in outburst in 2010. This source, hereafter Terzan 5 X-2, was only in outburst for $\sim$11 weeks and it was the first source that had a short accretion outburst (i.e., a duration of months instead of years) after which crust cooling was observed (\cite[Degenaar et al. 2011]{degenaar2011}). Surprisingly, the crust remained hot $>$4 years after the end of the outburst although it only went through a short accretion phase. The large observed temperatures and the fact that the crust was not able to get rid of the heat on a faster timescale indicate that the source might have unusual crust properties (\cite[Degenaar et al. 2013]{degenaar2013}). This source stands out from other crust cooling sources studied so far for two additional reasons. Firstly, its magnetic field strength is estimated to be $\sim10^{9}$ G, which is an order of magnitude larger than for other crust cooling neutron stars in LMXBs (see, e.g., Table 1 of \cite[Degenaar et al. 2015]{degenaar2015} and references therein). Secondly, Terzan 5 X-2 has a spin frequency of 11 Hz, whereas the typical observed spin frequency for other crust cooling sources is $\sim$500 Hz (\cite[Degenaar et al. 2015]{degenaar2015}). Because of its low spin frequency and relatively large magnetic field, it has been suggested that Terzan 5 X-2 has entered its spin-up phase towards an accreting millisecond pulsar more recently and thus may have started accreting more recently (e.g., see the discussion in \cite[Patruno et al. 2012]{2012patruno}). \\

\begin{figure}[h!]
	\vspace*{-0.25 cm}
	\begin{center}
		\includegraphics[width=\linewidth]{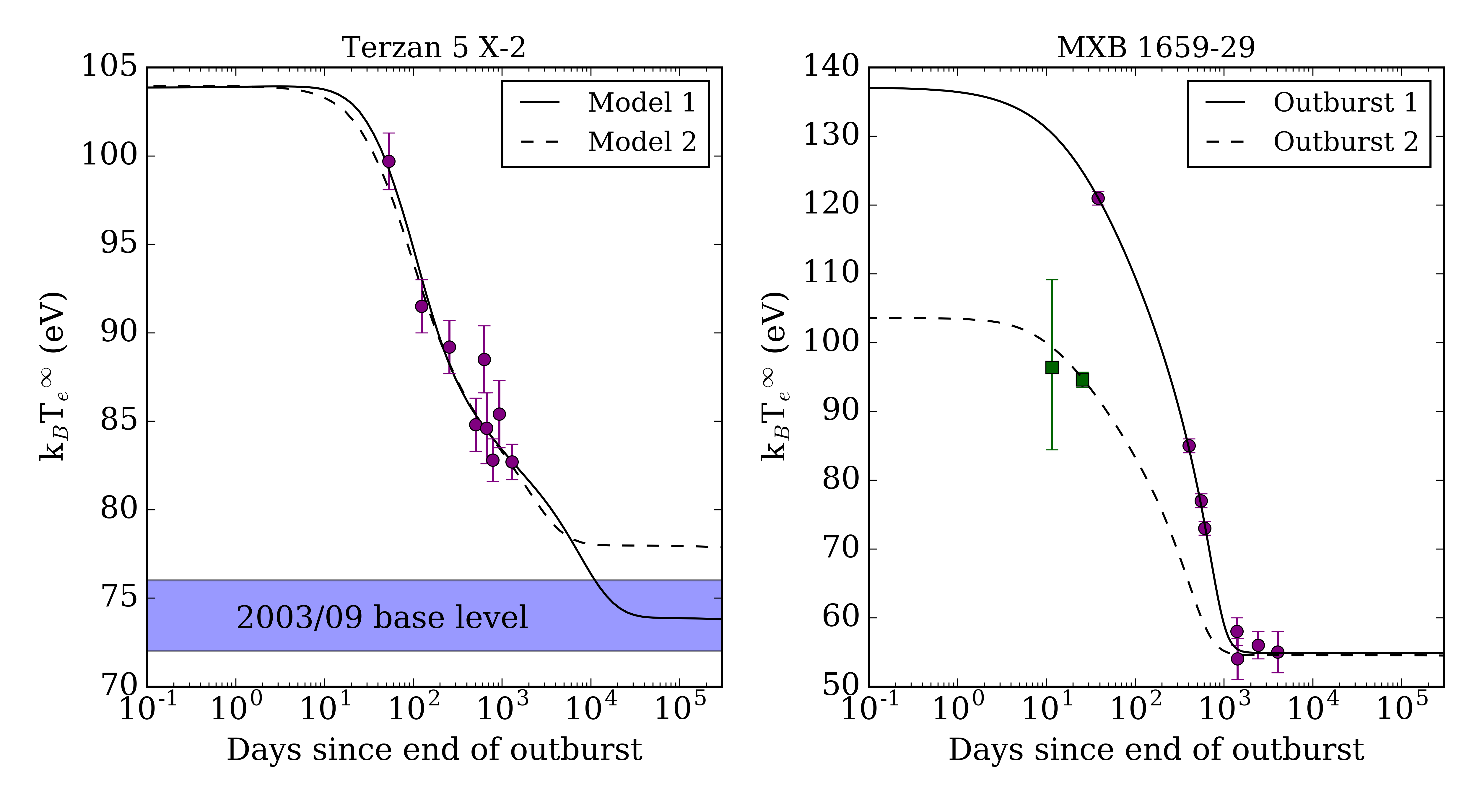} \bibitem[Degenaar \etal\ (2013)]{degenaar2013} \bibitem[Degenaar \etal\ (2015)]{degenaar2015}
		\vspace*{-0.7 cm}
		\caption{\textit{Left:} The calculated cooling curves for the two models that best describe the observed temperatures of Terzan 5 X-2 (data from \cite[Degenaar et al. 2013, 2015]{degenaar2013}). Model 1 includes a low conductivity region (with a large amount of impurities) in the crust and cools back to the pre-outburst quiescence temperature (indicated by the shaded region). Model 2 corresponds to a change in envelope composition after the accretion phase. \textit{Right:} Calculated cooling curve for MXB 1659-29 for outburst 1 (solid; using the data points described in \cite[Cackett et al. 2008, 2013]{Cackett2008}) and predicted cooling curve for outburst 2 (dashed; using the first two data points of \cite[Parikh et al., 2018)]{2018parikh}. The only difference between the model parameters is the shallow heating strength.}
		\label{fig1}
	\end{center}
\end{figure}

Using the cooling data points presented by \cite[Degenaar et al. (2013)]{degenaar2013} and \cite[Degenaar et al. (2015)]{degenaar2015} we have calculated cooling models that can explain the observed thermal behaviour of Terzan 5 X-2. We also use the temperatures obtained before the outburst by the same authors to constrain the temperature where the crust and core are in thermal equilibrium (base level). We provide two scenario's that are consistent with the observations as shown in the left panel of Figure \ref{fig1}. Firstly, we can explain that the crust is still hot compared to its observed pre-outburst temperature if the crust has a low thermal conductivity layer in the neutron drip region ($\rho$ = 4 $\times$ 10$^{11}$ - 8~$\times$~10$^{13}$~g~cm$^{-3}$) due to a large amount of impurities in the nuclear lattice (model 1). It was already found for XTE J1701-462 that the amount of impurities can be larger in this region of the crust (\cite[Page \& Reddy 2013]{page2013}). The presence of impurities in the atomic lattice can significantly increase the scattering rate of the heat carriers and consequently reduce the thermal conductivity of the crust (\cite[Yakovlev \& Urpin 1980]{1980yakovlev}). The amount of impurities that must be present in this scenario for Terzan 5 X-2 is over 10 times larger than what has been inferred from the other cooling sources. This means that in this region, the crust might be more amorphous and unstructured than previously found. We also obtain an alternative scenario, in which the composition of the envelope has changed (contains more light elements) compared to the pre-outburst situation (model 2). Consequently, the observed base temperature after the outburst appears larger than the observed pre-outburst temperature although the interior temperature did not change (\cite[Brown et al. 2002]{Brown2002}). While it is possible to distinguish between the provided scenarios with future observations, this might only be achievable in $\sim$15-20 years from now since only then the models predict significantly differences in the cooling curves (as is clear from the left panel of Figure 1).

%
\firstsection
\vspace*{-0.3 cm}
\section{MXB 1659-29: On the origin of shallow heating}


MXB 1659-298 was the second source for which a cooling crust was observed (after its 1999-2001 outburst that lasted $\sim$2.5 years). In 2015 the source went back in outburst but this time it was only active for $\sim$1.8 years (\cite[Parikh et al. 2017a]{2017parikha}). This gives the rare opportunity to study crust cooling after multiple outbursts in the same source. The cooling after the MXB 1659-29 outburst ending in 2001 has been the subject of many observational and theoretical studies that have led to a better understanding of the properties of the neutron star interior (e.g., \cite[Wijnands et al. 2004]{2004wijnands}, \cite[Brown \& Cumming 2009]{Brown2009}, \cite[Page \& Reddy 2013]{page2013}, \cite[Deibel et al. 2017]{deibel2017}, \cite[Cumming et al. 2017]{Cumming2017}). Using the \textit{NSCool} code, we modelled the thermal evolution of the source after its first outburst and calculated predictions for the cooling curve that should be observable after the second outburst (see right panel of Figure \ref{fig1}). \\

We use the first two cooling data points obtained by \cite[Parikh et al. (2018)]{2018parikh} after the second outburst to constrain the shallow heating strength and depth during this outburst and provide predictions for the ongoing cooling. The right panel of Figure 1 shows the calculated cooling after both outbursts using the same physical parameters except for the shallow heating properties. We find that for these curves, the shallow heating is $\sim$3.2 MeV/nucleon and $\sim$1.5 MeV/nucleon during outburst 1 and outburst 2, respectively. The result that the shallow heating strength can be different between outbursts in the same source is in accordance with the results of \cite[Parikh et al. (2017b)]{2017parikhb}, \cite[Ootes et al. (2018)]{2018lsootes} and \cite[Deibel et al. (2015)]{deibel2015} and has important consequences for the mechanism behind shallow heating. For example, it is therefore unlikely that the shallow heating could be explained by uncertainties in the energy released by nuclear reactions in the crust, which are assumed to be proportional to the accretion rate (e.g., \cite[Horowitz et al. 2008]{horowitz2008}, \cite[Estrad\'e et al. 2011]{estrade2011}). \\

We find that the ratio of the shallow heating strength and total accreted mass are strikingly similar for both outbursts of MXB 1659-29 as was also found for the multiple outbursts of MAXI J0556-332 (\cite[Parikh et al. 2017b]{2018parikhb}). This indicates that the shallow heating mechanism might be related to properties of the accretion outburst, which are known to vary between outbursts (e.g., in duration, brightness, accretion rate, geometry). Multiple outbursts in the same system are needed to further test this hypothesis. Several additional observations of MXB 1659-29 during its second cooling phase will be performed, which will allow us to conclusively determine if only the properties of the shallow heating differ between the two cooling curves, or that other differences (e.g., the composition of the envelope) are also needed to explain both curves.

\end{document}